\documentclass[pra,twocolumn,aps]{revtex4}
\usepackage{amsmath}
\usepackage[dvips]{graphicx}

\begin{document}

\title{Control of decoherence in the generation of photon pairs from atomic ensembles}

\author{D. Felinto, C.~W. Chou, H. de Riedmatten, S.~V. Polyakov, and H.~J. Kimble}

\affiliation{Norman Bridge Laboratory of Physics 12-33,  California Institute of Technology, Pasadena, California 91125, USA}
\date{\today}

\begin{abstract}
We report an investigation to establish the physical mechanisms responsible for decoherence in the generation of photon pairs from atomic ensembles, via the protocol of Duan {\it et. al.} for long distance quantum communication [Nature (London) {\bf 414}, 413 (2001)] and present the experimental techniques necessary to properly control the process. We develop a theory to model in detail the decoherence process in experiments with magneto-optical traps. The inhomogeneous broadening of the ground state by the trap magnetic field is identified as the principal mechanism for decoherence. The theory includes the Zeeman structure of the atomic hyperfine levels used in the experiment, and the polarization of both excitation fields and detected photons. In conjunction with our theoretical analysis, we report a series of measurements to characterize and control the coherence time in our experimental setup. We use copropagating stimulated Raman spectroscopy to access directly the ground state energy distribution of the ensemble. These spectroscopic measurements allow us to switch off the trap magnetic field in a controlled way, optimizing the repetition rate for single-photon measurements. With the magnetic field off, we then measure nonclassical correlations for pairs of photons generated by the ensemble as a function of the storage time of the single collective atomic excitation. We report coherence times longer than 10 $\mu s$, corresponding to an increase of two orders of magnitude compared to previous results in cold ensembles. The coherence time is now two orders of magnitude longer than the duration of the excitation pulses. The comparison between these experimental results and the theory shows good agreement. Finally, we employ our theory to devise ways to improve the experiment by optical pumping to specific initial states.  
\end{abstract}

\maketitle

\section{Introduction}

Quantum memory is a key resource for many quantum-information protocols. Usually it is associated with the basic requirements for quantum computation~\cite{QI,QC}, but in recent years also quantum communication protocols started to rely on it. The requirement of memory was introduced in quantum communication as part of the idea for quantum repeaters~\cite{repeater_01,repeater_02}, a possible solution for the problem of quantum communication over long distances. In this case, memory is essential to increase the probability of success of the chain of conditional steps that underlies the protocol, and makes feasible scalable quantum networks.

A significant step toward the realization of the quantum repeater idea was a proposal by Duan, Lukin, Cirac, and Zoller (DLCZ) for its implementation using linear optics and atomic ensembles~\cite{DLCZ_01}. The DLCZ protocol is based on the generation of single photons  by spontaneous Raman scattering in atomic ensembles ~\cite{DLCZ_02}. The detection of a single photon in the forward propagating mode heralds the presence of a single collective atomic excitation in the sample, due to a collective enhancement effect. This excitation can be stored for a time up to the coherence time of the ground states of the atoms and then converted back into a light field. Entanglement of distant ensembles in the excitation number basis is generated by interference~\cite{EntInterference}, and extended to longer distances by entanglement swapping~\cite{teleportation_01,swap_01}. The final pairs of ensembles, far apart, can then be used for entanglement-based quantum cryptography~\cite{DLCZ_01,Ekert}, probabilistic quantum teleportation and violation of Bell inequality. This proposal has received much attention in the past two years and several groups are
presently pursuing its experimental implementation~\cite{DLCZ_03,DLCZ_06,DLCZ_08,DLCZ_04,DLCZ_05,kuzmich2004,DLCZ_07,DLCZ_harris}.

In this article, we analyze the decoherence processes present in the DLCZ protocol, and describe experiments to mitigate the problem.  We construct a theory for the decoherence process in the photon-pair generation. Particularly, our analysis concentrates in its implementation with cold atomic ensembles, but many results should also apply to studies with room-temperature ensembles in vapor cells. We propose various strategies to increase the system's coherence time, and introduce experimental techniques necessary for its characterization and control. We also report the first experimental steps in this direction, with an increase of more than two order of magnitude in the coherence time with respect to the previously reported works with cold atoms~\cite{DLCZ_03,DLCZ_04,DLCZ_05,kuzmich2004,DLCZ_harris}.

The coherence times reported up to now by the several groups working on the implementation of the protocol are all shorter or of the order of a couple of microseconds. Furthermore, for all experiments to date, the reported coherence times are of the order of the excitation pulses duration. However, for using this system as a quantum memory, it is important to obtain storage time much longer than the excitation pulses. Moreover, for the DLCZ protocol to become a viable alternative for long distance quantum communication, long coherence time is crucial and major efforts are required to increase it. The main goal of the present article is then to provide the initial steps in this direction, and to establish several techniques and ideas for the next steps.

Only two types of systems have been employed in the experiments up to now: vapor cells~\cite{DLCZ_06,DLCZ_08,DLCZ_07} and cold atoms in magneto-optical traps~\cite{DLCZ_03,DLCZ_04,DLCZ_05,kuzmich2004,DLCZ_harris}. In both systems, however, the experiments have not achieved yet their respective state-of-the-art coherence times. The vapor-cell studies, for example, did not employ paraffin coated cells~\cite{parafin_01,parafin_02}; the coherence times were effectively limited to the time the atoms take to diffuse out of the excitation region, which is of the order of microseconds. Recently, high fidelity atomic quantum memory of the state of a light pulse was achieved with such paraffin coated cells~\cite{quant_mem} with memory times of up to 4~ms. Coherence times of tens of milliseconds, however, are commonly achieved in this system~\cite{parafin_03}, and there are reports of coherence times as high as one second~\cite{parafin_01}. The difference in these values is largely due to measurements of decay of different coherent processes~\cite{parafin_01}. How the coherence required for the generation of photon pairs from atomic vapors will decay as the atoms collide with the walls of paraffin coated cells is still to be determined.

The use of atomic traps to generate photon pairs for the DLCZ protocol has the advantage of providing a high density of atoms distributed in a small spectral region, due to the suppression of Doppler broadening by the cooling process. This allows the use of excitation laser pulses tuned closer to resonance, which requires much less power and makes it easier to filter the excitation pulses from the Ramam-scattered photons. However, atomic traps also introduce a different set of complications. In the case of the magneto-optical traps (MOT) used up to now, the magnetic field of the trap induces decoherence on a timescale of the order or smaller than a few hundreds nanoseconds~\cite{kuzmich2004,DLCZ_05,DLCZ_harris}. The first results with the MOT magnetic field off are reported in the present article, with coherence times on the order of 10~$\mu$s. As will be discussed below in detail, a better nulling of the magnetic field combined with optical pumping to specific Zeeman levels might increase the coherence time, in a straightforward way, to hundreds of microseconds.

Further improvements with MOTs would face the problem of diffusion of atoms from the excitation region and, most troublesome, from the MOT itself. This problem can in principle be mitigated by improved cooling techniques. However, along these lines, it would be difficult to increase the coherence time above a couple of milliseconds. A possible solution then is to use an optical dipole trap to hold the atoms during the write-and-read process. Hyperfine coherence times of hundreds of milliseconds have already been observed in such traps~\cite{dipole01,dipole02}.

In the following, Secs.~\ref{theory} and~\ref{proposals} are devoted to theoretical results and Sec.~\ref{experiments} to associated experiments. In Sec.~\ref{DLCZ} we give a general introduction to the photon-pair generation process behind the DLCZ protocol. In Sec.~\ref{decoherence}, we derive a theory for the probability of joint detection of these photons pairs generated from an atomic ensemble in a magneto-optical trap. This theory is a direct extension of a previous theoretical treatment reported in Ref.~\onlinecite{DLCZ_02}, to which we added explicitly the reading process and the Zeeman structure of the levels. In this way, we are able to model the action of the magnetic field over the atoms, and to study the dependence of the correlations with the light polarization.

Section~\ref{experiments} describes an experimental investigation leading to the nulling of the magnetic field in the photon-pair correlation measurements, with the subsequent increase in the system coherence time and degree of correlation. In Sec.~\ref{null}, we describe a series of Raman-spectroscopy experiments to characterize the system and optimize the process of zeroing the magnetic field. We determine the set of experimental conditions that result in a good compromise between atomic density and magnetic field cancellation, which we used in the correlation measurements. Section~\ref{correlation measurements} describes then measurements of nonclassical correlations for the photon pairs generated by the MOT. We compare results with magnetic field on and with magnetic field off. The magnetic field off measurements present a higher degree of correlation, and a hundred times larger coherence time. We compare the shape of the experimental curves with magnetic field on and off to our theory, obtaining good agreement. We also show how the two-photon wavepacket that describes the detailed temporal structure of the photon pair generation is modified by the magnetic field.

Finally, based on the procedure for comparison between theory and experiment described in Sec.~\ref{correlation measurements}, we formulate in Sec.~\ref{proposals} a proposal to improve our experimental signal. We suggest using a combination of optical pumping to a specific initial state and polarization of the light fields to increase both our detection efficiency and coherence time. Section~\ref{conclusion} is dedicated to our conclusions.

\section{Theory}
\label{theory}

The basic theory for the DLCZ protocol is described in Refs.~\cite{DLCZ_01} and~\cite{DLCZ_02}. The general idea of the protocol is treated in Ref.~\cite{DLCZ_01}, while Ref.~\cite{DLCZ_02} gives a detailed analysis of the collective emission of photons through spontaneous Raman scattering following excitation by free-space light. Section~\ref{decoherence} provides an extension of the theoretical treatment of Ref.~\cite{DLCZ_02} to better account for our experimental conditions. The emphasis here is the modeling of the decoherence process due to external magnetic fields, and in particular for experiments using magneto-optical traps. To model this decoherence, the essential elements to be introduced in the previous theory of Ref.~\cite{DLCZ_02} are the Zeeman structure of all levels and an explicit treatment of the reading process. On the other hand, the theory in Sec.~\ref{decoherence} is a simplification of the treatment of Ref.~\cite{DLCZ_02} concerning the spatial mode of the photons. We consider only the forward, collectively enhanced emission. The reading process is also treated in a simplified, perturbative way, while the experiments are done with stronger read pulses on resonance. This later difference between theory and experiment will result in some noticeable discrepancy in Sec.~\ref{2photon}, where we discuss measurements of the two-photon wavepacket of the pair-generation process. In general, however, the comparison between theory and experiment performed in Sec.~\ref{correlation measurements} results in very good agreement, which indicates that the theory in Sec.~\ref{decoherence} takes into account the essential physical elements behind the decoherence process.

\subsection{Photon pair generation}
\label{DLCZ}

\begin{figure}[th]
\centerline{\includegraphics[width = 8.5cm]{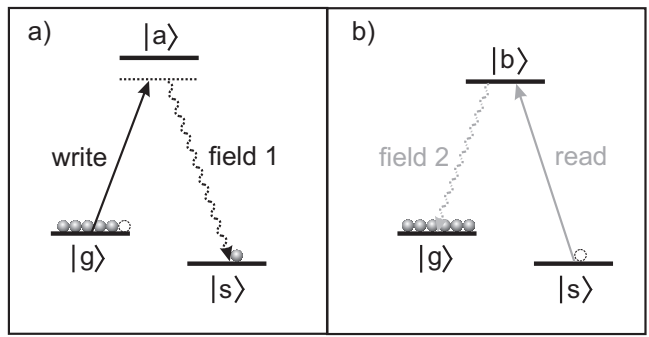}}
\caption{Relevant level structure of the atoms in the ensemble for (a) writing and (b) reading processes, with $|g\rangle$ the initial ground state and $|s\rangle$ the ground state for storing an excitation. $|a\rangle$ and $|b \rangle$ are excited states. The transition $|g\rangle \rightarrow |a\rangle$ is initially coupled by a classical laser pulse (write beam) detuned from resonance, and the forward-scattered Stokes light (field 1) comes from the transition $|a\rangle \rightarrow |s\rangle$, which has different polarization or frequency to the write light. A classical read
pulse then couples the transition $|s\rangle \rightarrow |b\rangle$, leading to the emission of forward-scattered
anti-Stokes light (field 2) from the transition $|b\rangle \rightarrow |g\rangle$. \label{3level}}
\end{figure}

The building block of the DLCZ protocol is an ensemble of $N$ identical atoms with lambda-type energy level configuration as shown in Fig.~\ref{3level}, which we briefly discuss here in an ideal setting. In the experiments discussed in this article, the lower states $|g\rangle$ and $|s\rangle$ are hyperfine sublevels of the electronic ground state of Cesium atoms. First, all atoms are prepared in the state $|g\rangle$. By sending in a weak, off-resonant laser pulse, one atom of the ensemble might be transfered from $|g\rangle$ to $|s\rangle$, thus emitting a photon (field $1$) at a frequency or polarization different from the original exciting field. A key element of the protocol is the collective enhancement of this spontaneous Raman scattering in a forward direction, which is determined by the spatial mode of the laser pulse and the geometry of the excitation region~\cite{DLCZ_02}. If the laser intensity is low enough so that two excitations are very unlikely, the detection of the photon generated in this process is a signature that the ensemble was excited to a symmetrical collective state~\cite{DLCZ_01,DLCZ_02}, which in the ideal case can be explicitly written as 
\begin{equation}
|1_{a}\rangle=\frac{1}{\sqrt{N}}\sum^{N}_{i=1}|g\rangle_{1}\cdots|s\rangle_{i}\cdots|g\rangle_{N}
\label{symmetrized}\text{,}
\end{equation}
where the sum goes over all atoms addressed by the laser pulse, and $|1_{a}\rangle$ indicates the state of the atomic ensemble with just one excitation. This is the ``writing" step of the protocol (Fig.~\ref{3level}a).

Since the excitation probability $\chi$ is very small, the whole state of the system consisting of atoms and forward-scattered mode of light is in the following form:
\begin{equation}
|\phi\rangle =|0_{a}\rangle|0_{1}\rangle+e^{i\beta}\sqrt{\chi}\,|1_{a}\rangle|1_{1}\rangle+O(\chi)\text{,}
\label{scas}
\end{equation}
where $\chi<<1$, $|n_{1}\rangle$ stands for the state of the forward-propagating light field $1$ with $n$ photons, $\beta$ is a phase set by propagation to and from the ensemble, and $|0_{a}\rangle\equiv\bigotimes_{i}^{N_a}|g\rangle_{i}$. $O(\chi)$ represents all the other possible excitation processes, which in the ideal case occur with probabilities of order $\chi^{2}$. The system remains in this state for a time on the order of the lifetime of the ground states. By sending in a second (``read") pulse resonant with the $|s\rangle \rightarrow |b\rangle$ transition, the state of the atomic ensemble can be transferred deterministically (read out) to another forward-propagating light field 2 at the $|b\rangle \rightarrow |g\rangle$ transition (see Fig.~\ref{3level}b).  In this way, it is possible to access the quantum state of the atoms. This reading process is then closely related to low-light-level Electromagnetically Induced Transparency~\cite{LukinReview,SHarris02}. After the read out, the state of the system becomes:
\begin{equation}
|\phi\rangle =|0_{1}\rangle|0_{2}\rangle+e^{i\gamma}\sqrt{\chi}\,|1_{1}\rangle|1_{2}\rangle+O(\chi)\text{,}
\label{scas2}
\end{equation}
where $\gamma$ is a phase that includes $\beta$ and the propagation phases to and from the ensemble related to the reading process. Fields 1 and 2 exhibit now strong correlations in the photon number basis, and can be described as photon pairs. These non-classical correlations can be measured by photoelectric detection. Since the field 2 maps the state of the atoms, the correlations between field 1 and field 2 can then be used to infer correlations between field 1 and the collective atomic excitations in the sample.
\subsection{Decoherence}
\label{decoherence}

In order to analyze the decoherence process in the generation of pairs from an atomic ensemble as described in Sec.~\ref{DLCZ}, we need to expand the theoretical treatment of Ref.~\cite{DLCZ_02} to include other experimentally relevant features. For our experiments in particular, it is essential to include the  splitting of the Zeeman structure of the atomic ground states due to the magnetic field. The MOT quadrupole field generates an inhomogeneous distribution of splittings throughout the ensemble. As the system evolves in time, this results in dephasing between different regions of the atomic cloud, and in a respective decay of the coherence of the collective state. It is also important to include explicitly the
reading process in the theory. For simplicity, this is done by considering a read process similar to the write process, i.e., with small probability of excitation and detuned from the excited state. Note that in the actual experiment, the read beam is stronger than the write beam and is on resonance. This will lead to small discrepancies when comparing the experimental results to the theory, that will be discussed in section \ref{2photon}. 

The inclusion of Zeeman structure in the theory allows a detailed discussion of the effect of light polarization in the experiment. This is important to evaluate different excitation and detection schemes. It also gives a better description of the initial state, and of its role on the subsequent coherent pair generation. Together, the analysis of different polarization schemes and of different initial states led to specific proposals of ways to improve the whole process. These features of the theory are not specifically related to the MOT magnetic field, and should apply to pair generation in other systems, like vapor cells or dipole traps.

\begin{figure}[htb]
    \begin{center}
        \includegraphics[width = 6.0cm,angle=270]{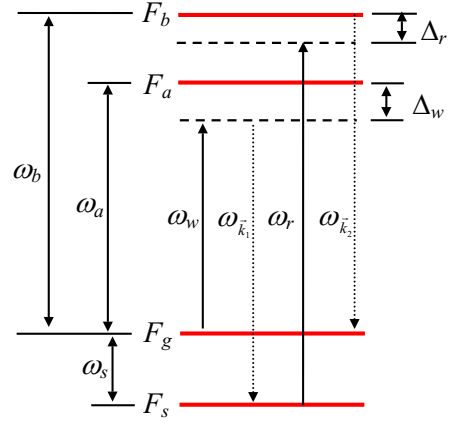}
    \end{center}
 
    \caption{Energy level scheme considered for the atomic ensembles}
    \label{fig:Fig1}
\end{figure}

Our treatment starts by considering a sample of $N$ four-level atoms, such as in Fig.~\ref{fig:Fig1}. The four levels represent manifolds of Zeeman sublevels and are indicated by their respective $F$ quantum numbers. A specific state of the $F_j$ manifold of the $i$-th atom is represented by its ket $|m_j\rangle_i$, where $m_j$ is the azimuthal quantum number. Two pumping fields act on the system, namely a write field $\vec{{\cal E}}_{ga}$ and a read field $\vec{{\cal E}}_{sb}$, where 
\begin{subequations}
\begin{eqnarray}
\vec{{\cal E}}_{ga}(\vec{r},t) &=& u_w(\vec{r},t)e^{i(k_wz-\omega_wt)}\vec{e}_{p_w} \,, \\ [0.5cm]
\vec{{\cal E}}_{sb}(\vec{r},t) &=& u_r(\vec{r},t)e^{i(k_rz-\omega_rt)}\vec{e}_{p_r} \,,
\end{eqnarray}
\end{subequations}
which couple the transitions $F_g \rightarrow F_a$ and $F_s \rightarrow F_b$, respectively. The functions $u_w$ and $u_r$ give the slowly-varying envelopes of the {\it write} and {\it read} pulses, respectively, and $\vec{e}_{p_w}$ and $\vec{e}_{p_r}$ are their polarization vectors. As a result of their action, two Raman fields are spontaneously generated in the sample:
\begin{subequations}
\begin{eqnarray}
\hat{\vec{{\cal E}}}_{sa}(\vec{r},t) &\propto& \sum_{p_1} \int d\vec{k}_1 \hat{a}_{\vec{k}_1p_1}e^{i(\vec{k}_1\cdot \,\vec{r} - \omega_{\vec{k}_1}t)} \vec{e}_{p_1} \,, \\ [0.5cm]
\hat{\vec{{\cal E}}}_{gb}(\vec{r},t) &\propto& \sum_{p_2} \int d\vec{k}_2 \hat{b}_{\vec{k}_2p_2}e^{i(\vec{k}_2\cdot \,\vec{r} - \omega_{\vec{k}_2}t)} \vec{e}_{p_2} \,,
\end{eqnarray}
\end{subequations}
where  $\omega_{\vec{k}_i} = |\vec{k}_i|c$ and $p_i$ is a label for the field polarization. $\hat{a}_{\vec{k}_1p_1}$ and
$\hat{b}_{\vec{k}_2p_2}$ are the annihilation operators for the Raman fields 1 and 2, respectively, which couple the transitions $F_s \rightarrow F_a$ and $F_g \rightarrow F_b$. The state of field 1 with just one photon excited in mode $\vec{k}_1 p_1$ will be designated by $|1_{\vec{k}_1p_1}\rangle$. A similar notation will be used for field 2.

The Hamiltonian for the system of $N$ atoms can be written as
\begin{equation}
\hat{H}(t) = \hat{H}_0 + \hat{V}(t)\,,
\end{equation}
where
\begin{align}
\hat{H}_0 &= \sum_{i=1}^N \Bigg\{ \sum_{m_s = -F_s}^{F_s} \left( -\hbar \omega_s + \mu_B g_s m_s B_{z_i}\right) |m_s\rangle_i \langle m_s| \nonumber \\
& + \sum_{m_g = -F_g}^{F_g} \mu_B g_g m_g B_{z_i} |m_g\rangle_i \langle m_g| \nonumber \\
& + \sum_{m_a = -F_a}^{F_a} \hbar \omega_a |m_a\rangle_i \langle m_a| + \sum_{m_b = -F_b}^{F_b} \hbar \omega_b |m_b\rangle_i \langle m_b| \Bigg\} \,
\end{align}
is the free-atom Hamiltonian, and
\begin{align}
\hat{V}(t) &= \sum_{i=1}^N \Bigg\{ \sum_{m_a = -F_a}^{F_a} \sum_{m_g = -F_g}^{F_g} \left( -\vec{d}_{m_a m_g} \cdot \vec{{\cal E}}_{ga} \right)|m_a\rangle_i \langle m_g| \nonumber \\
& + \sum_{m_s = -F_s}^{F_s} \sum_{m_a = -F_a}^{F_a} \left( -\vec{d}_{m_s m_a} \cdot \hat{\vec{{\cal E}}}_{sa}^{\dagger} \right) |m_s\rangle_i \langle m_a| \nonumber \\
& + \sum_{m_b = -F_b}^{F_b} \sum_{m_s = -F_s}^{F_s} \left( -\vec{d}_{m_b m_s} \cdot \vec{{\cal E}}_{sb} \right) |m_b\rangle_i \langle m_s| \nonumber \\
& + \sum_{m_g = -F_g}^{F_g} \sum_{m_b = -F_b}^{F_b} \left( -\vec{d}_{m_g m_b} \cdot \hat{\vec{{\cal E}}}_{gb}^{\dagger} \right) |m_g\rangle_i \langle m_b|\Bigg\} \,
\end{align}
gives the time-dependent interaction Hamiltonian. $\vec{d}_{jk}$ is the dipole moment for the $j \rightarrow k$ transition, $\mu_B$ the Bohr magneton, $g_j$ the hyperfine Land\'e factor for level $F_j$, and $B_{z_i}$ is the magnetic field in the position of the $i$-th atom. The magnetic field direction is taken as the quantization $z$ axis. We neglect the Zeeman splitting of the excited states since we want to investigate a situation where it is always smaller than the excited-states natural linewidths. The factors $-\vec{d}_{jk}\cdot\vec{{\cal E}}_{kj}$ can also be written as
\begin{subequations}
\begin{align}
-\vec{d}_{m_a m_g}\cdot\vec{{\cal E}}_{ga} &= K_{m_a m_g} u_w(\vec{r}_i,t) e^{i(k_w z_i - \omega_w t)} \,, \\
-\vec{d}_{m_s m_a}\cdot\hat{\vec{{\cal E}}}_{sa}^{\dagger} &= \sum_{p_1} \int d\vec{k}_1 K_{m_s m_a}^{\vec{k}_1 p_1} \hat{a}_{\vec{k}_1 p_1}^{\dagger} e^{-i(\vec{k}_1\cdot \,\vec{r} - \omega_{\vec{k}_1}t)} \,, \\
-\vec{d}_{m_b m_s}\cdot\vec{{\cal E}}_{sb} &= K_{m_b m_s} u_r(\vec{r}_i,t) e^{i(k_r z_i - \omega_r t)} \,, \\
-\vec{d}_{m_g m_b}\cdot\hat{\vec{{\cal E}}}_{gb}^{\dagger} &= \sum_{p_2} \int d\vec{k}_2 K_{m_g m_b}^{\vec{k}_2 p_2} \hat{b}_{\vec{k}_2 p_2}^{\dagger} e^{-i(\vec{k}_2\cdot \,\vec{r} - \omega_{\vec{k}_2}t)} \,,
\end{align}
\end{subequations}
where $K_{m_a m_g}$, $K_{m_s m_a}^{\vec{k}_1 p_1}$, $K_{m_b m_s}$, and $K_{m_g m_b}^{\vec{k}_2 p_2}$ are coupling constants for the corresponding transition.

The temporal evolution of the coupled system consisting of ensemble + Raman fields is described by the evolution of its density matrix $\hat{\rho}(t)$. In the interaction picture, the corresponding operator $\hat{\rho}_I(t)$ is given by
\begin{equation}
\hat{\rho}_I(t) = \hat{U}_I(t) \hat{\rho}(0) \hat{U}_I^{\dagger}(t)\,, \label{density1}
\end{equation}
where $\hat{U}_I(t)$ is the temporal evolution operator, and the initial state $\hat{\rho}(0)$ can be written as
\begin{equation}
\hat{\rho}(0) = \hat{\rho}_{F_1}(0)\otimes \hat{\rho}_{F_2}(0) \otimes \hat{\rho}_1(0) \otimes \hat{\rho}_2(0) \otimes \cdots \otimes \hat{\rho}_N(0) \,,
\end{equation}
with $\hat{\rho}_{F_1}(0)$ the initial state of field 1, $\hat{\rho}_{F_2}(0)$ the initial state of field 2, and $\hat{\rho}_i(0)$ the initial state of the $i$-th atom. For most of what follows, we will be interested in the case where the fields 1 and 2 are initially vacuum states, $\hat{\rho}_{F_1}(0) = |vac_{F_1}\rangle\langle vac_{F_1}|$ and $\hat{\rho}_{F_2}(0) = |vac_{F_2}\rangle\langle vac_{F_2}|$, and all atoms are initially in the same incoherent distribution over the Zeeman sublevels of the $F_g$ state:
\begin{equation}
\hat{\rho}_i(0) = \sum_{m_g = -F_g}^{F_g} D_{m_g} |m_g\rangle_i\langle m_g|\,,
\label{distribution}
\end{equation}
with $D_{m_g}$ giving the probability of finding an atom in the $m_g$ state at $t=0$. In section \ref{proposals} however, we will consider the case where all the atoms are optically pumped in one of the Zeeman sublevel ($m_F=0$).

The operator $\hat{U}(t)$ can be written as a Dyson series in the form
\begin{equation}
\hat{U}_I(t) = 1 + \sum_{i=1}^N \hat{{\cal U}}_i^{(1)}(t) + \sum_{i=1}^N \hat{{\cal U}}_i^{(2)}(t) + \cdots \,, \label{Dyson2}
\end{equation}
where
\begin{eqnarray}
\hat{{\cal U}}_i^{(1)}(t) &=& \left( -\frac{i}{\hbar} \right) \int_0^t dt^{\prime}\hat{{\cal V}}_i(t^{\prime}) \,, \nonumber \\ [0.3cm]
\hat{{\cal U}}_i^{(2)}(t) &=& \left( -\frac{i}{\hbar} \right)^2 \int_0^t dt^{\prime} \int_0^{t^{\prime}} dt^{\prime\prime}\hat{{\cal V}}_i(t^{\prime})\hat{{\cal V}}_i(t^{\prime\prime}) \,,
\end{eqnarray}
and so on. The single-atom interaction operator $\hat{{\cal V}}_i(t)$ is defined from the expression for the general interaction Hamiltonian $\hat{V}_I(t)$ in the interaction picture as
\begin{equation}
\hat{V}_I(t) = e^{i\hat{H}_0 t/\hbar} \hat{V}(t) e^{-i\hat{H}_0 t/\hbar} = \sum_{i=1}^N \hat{{\cal V}}_i(t) \,.
\end{equation}

\subsubsection{Probability for joint detections}
\label{section_p1}

We want to calculate in the lowest order of perturbation the probability of detecting a single photon in field 1 followed by another photon in field 2. The first step is then to calculate the restriction of the coupled state $\hat{\rho}(t)$ to the space of states of fields 1 and 2:
\begin{equation}
\hat{\rho}_{F_1 F_2}(t) = \mbox{\large Tr}_A \left[ \hat{\rho}(t) \right] \,.
\end{equation}
The symbol $\mbox{\large Tr}_A$ indicates a partial trace over all atomic states. The probability for detecting two photons, one in mode $\vec{k}_1p_1$ and the other in mode $\vec{k}_2p_2$, up to time $t$ is then given by
\begin{align}
p_{12}^{th}(t,\vec{k}_1p_1,\vec{k}_2p_2) &= \langle 1_{\vec{k}_1p_1}| \langle 1_{\vec{k}_2p_2}|\hat{\rho}_{F_1 F_2}(t) |1_{\vec{k}_2p_2}\rangle|1_{\vec{k}_1p_1}\rangle \nonumber \\ &= \langle 1_{\vec{k}_1p_1}| \langle 1_{\vec{k}_2p_2}|\mbox{\large Tr}_A \left[ \hat{\rho}(t) \right] |1_{\vec{k}_2p_2}\rangle|1_{\vec{k}_1p_1}\rangle  . \label{p12th}\nonumber \\
\end{align}
Since all atoms are initially in the ground state $F_g$, the lowest order term of series~\eqref{Dyson2} that results in a single photon in field 1 and another photon in field 2 is the fifth term, which accounts for the four transitions carried successively by the write field, photon 1, read field, and photon 2, respectively. Substituting Eqs.~\eqref{density1} and~\eqref{Dyson2} into Eq.~\eqref{p12th} and keeping only the lowest order term, we arrive then at
\begin{align}
& p_{12}^{th}(t,\vec{k}_1p_1,\vec{k}_2p_2) = \nonumber \\ & \sum_{i,j = 1}^N \langle 1_{\vec{k}_1p_1}| \langle 1_{\vec{k}_2p_2}| \mbox{\large Tr}_A \negthickspace\left[ \hat{{\cal U}}_{i}^{(4)}(t) \hat{\rho}(0) \hat{{\cal U}}_{j}^{(4)\dagger}(t) \right]\negthickspace|1_{\vec{k}_2p_2}\rangle |1_{\vec{k}_1p_1}\rangle . \nonumber \\ \label{p1_2}
\end{align}
Note that $\hat{{\cal U}}_{k}^{(4)}$ acts only over the $k$-th atom. Thus, the trace $\mbox{\large Tr}_A$ on each term of the double sum can be written as a trace $\mbox{\large Tr}_k$ over the states of the atoms at which the $\hat{{\cal U}}_{k}^{(4)}$ operator is acting, since all other atoms remain in their initial state. Two different cases are present in Eq.~\eqref{p1_2}. If $i \neq j$, the two operators act over two different atoms and the initial state $\hat{\rho}(0)$ simplifies to $\hat{\rho}_{F_1}(0)\otimes\hat{\rho}_{F_2}(0)\otimes\hat{\rho}_i(0)\otimes\hat{\rho}_j(0)$. If $i = j$, then $\hat{\rho}(0) \rightarrow \hat{\rho}_{F_1}(0)\otimes\hat{\rho}_{F_2}(0)\otimes\hat{\rho}_i(0)$. With these observations in mind, we see that Eq.~\eqref{p1_2} can then be written as
\begin{align}
& p_{12}^{th}(t,\vec{k}_1p_1,\vec{k}_2p_2) = \nonumber \\
& \underset{i\neq j}{\sum_{i,j = 1}^N} \langle 1_{\vec{k}_1p_1}| \langle 1_{\vec{k}_2p_2}|\mbox{\large Tr}_i \left[ \hat{{\cal U}}_{i}^{(4)}(t) \hat{\rho}_i(0)\right]|vac_{F_2}\rangle |vac_{F_1}\rangle \nonumber \\
 & \quad\qquad \times \langle vac_{F_1}|\langle vac_{F_2}|\mbox{\large Tr}_j\left[\hat{\rho}_j(0)\hat{{\cal U}}_{j}^{(4)\dagger}(t) \right]|1_{\vec{k}_2p_2}\rangle |1_{\vec{k}_1p_1}\rangle \nonumber \\
 & + \sum_{i = 1}^N \langle 1_{\vec{k}_1p_1}| \langle 1_{\vec{k}_2p_2}| \mbox{\large Tr}_i \left[ \hat{{\cal U}}_{i}^{(4)}(t)\hat{\rho}_{F_1}(0) \otimes\hat{\rho}_{F_2}(0) \otimes \hat{\rho}_i(0) \right. \nonumber \\
 & \quad\qquad \times \left. \hat{{\cal U}}_{i}^{(4)\dagger}(t)\right]|1_{\vec{k}_2p_2}\rangle |1_{\vec{k}_1p_1}\rangle . \nonumber \\ \label{p12_3}
\end{align}
Substituting Eq.~\eqref{distribution}, we have
\begin{align}
& p_{12}^{th}(t,\vec{k}_1p_1,\vec{k}_2p_2) = \left| \sum_{i = 1}^N \sum_{m_g = -F_g}^{F_g} D_{mg} A_i(m_g,m_g) \right|^2 \nonumber \\
  & \qquad\qquad + \sum_{i = 1}^N \sum_{m_g^{\prime} = -F_g}^{F_g}\sum_{m_g = -F_g}^{F_g} D_{mg} |A_i(m_g^{\prime},m_g)|^2 \nonumber \\ & \qquad\qquad - \sum_{i = 1}^N \left|\sum_{m_g = -F_g}^{F_g} D_{mg} A_i(m_g,m_g)\right|^2 \,, \label{p12_4}
\end{align}
where
\begin{align}
A_i(&m_g^{\prime},m_g) = \nonumber \\&\langle 1_{\vec{k}_1p_1}| \langle 1_{\vec{k}_2p_2}| \langle m_g^{\prime}|_i \hat{{\cal U}}_{i}^{(4)}(t) |m_g\rangle_i |vac_{F_2}\rangle |vac_{F_1}\rangle. \label{Ai}
\end{align}

Note that the first term on the right side of Eq.~\eqref{p12_4} scales as $N^2$, while the two remaining terms scale with $N$ only. Since we are interested in the limit of large $N$, we can then approximate
\begin{align}
& p_{12}^{th}(t,\vec{k}_1p_1,\vec{k}_2p_2) = \left| \sum_{i = 1}^N \sum_{m_g = -F_g}^{F_g} D_{mg} A_i(m_g,m_g) \right|^2. \label{p12_5}
\end{align}
Thus, for large $N$, only transitions that start and end in the same state contribute to the pair generation. This result can be understood as a constructive interference between all pathways that connect the ensemble back to its initial state, after which it is not possible to distinguish which atom made the transition~\cite{vanEnk}. Pathways connecting different initial and final states leave a trace in the ensemble, which in principle can give information on which specific atom made the transition. In this last case, the number of possible pathways generating the pair of photons is then linearly proportional to the number of atoms $N$. Eq (\ref{p12_5}) expresses the collective enhancement that is essential to the scheme of ref. \cite{DLCZ_01}.

Finally, substituting the specific expressions for $\hat{{\cal U}}_{i}^{(4)}(t)$ and $\hat{{\cal V}}_i(t)$, we find that $A_i(m_g,m_g)$ can be written as
\begin{align}
A_i(m_g,m_g) &= \sum_{m_s = -F_s}^{F_s} \!\!\!\frac{d(m_g,ms)}{\hbar^4} \, e^{i(k_r z_i +k_wz_i - \vec{k}_1\cdot\vec{r}_i- \vec{k}_2\cdot\vec{r}_i)} \nonumber \\
& \times \int_0^t \!\!\! dt^{\prime} e^{i(\Delta \omega_{\vec{k}_2}-\Delta_r+a_{ig})t^{\prime}}  \nonumber \\
& \times \int_0^{t^{\prime}} \!\!\! dt^{\prime\prime}u_r(\vec{r}_i,t^{\prime\prime})e^{i(\Delta_r-a_{is})t^{\prime\prime}} \nonumber \\
& \times \int_0^{t^{\prime\prime}} \!\!\! dt^{\prime\prime\prime} e^{i(\Delta \omega_{\vec{k}_1}-\Delta_w+a_{is})t^{\prime\prime\prime}} \nonumber \\
& \times \int_0^{t^{\prime\prime\prime}} \!\!\! dt^{\prime \nu} u_w(\vec{r}_i,t^{\prime\nu})e^{i(\Delta_w-a_{ig})t^{\prime \nu}} \,, \label{Ai_2}
\end{align}
where $\Delta_w = \omega_a - \omega_w$, $\Delta_r = \omega_b + \omega_{s} - \omega_r$, $\Delta \omega_{\vec{k}_1} = \omega_{\vec{k_1}} - \omega_w - \omega_{s}$, $\Delta \omega_{\vec{k}_2} = \omega_{\vec{k_2}} - \omega_r + \omega_{s}$, and
\begin{equation}
d(m_g,m_s) = \!\!\sum_{m_b = -F_b}^{F_b} \sum_{m_a = -F_a}^{F_a} \!\!\!K_{m_g m_b}^{\vec{k}_2p_2} K_{m_b m_s}^{r} K_{m_s m_a}^{\vec{k}_1p_1} K_{m_a m_g}^{w}
\end{equation}
gives the strength of an specific excitation pathway in which the atom starts at $m_g$, then goes to $m_s$, and ends at $m_g$ again. The Zeeman splittings are written in terms of the parameters $a_{ig}=\mu_B g_g m_g B_{z_i}/\hbar$ and $a_{is} = \mu_B g_s m_s B_{z_i}/\hbar$.

\subsubsection{Forward emission}

In order to simplify the following analysis while keeping the essential trends of the temporal dynamics, we will focus now on the treatment of the forward, resonant emission from the atomic ensemble. In the forward direction, the light emitted by the sample satisfies the phase-matching condition
\begin{equation}
k_r z_i + k_w z_i - \vec{k}_1\cdot\vec{r}_i - \vec{k}_2 \cdot\vec{r}_i = 0.
\end{equation}
The resonant conditions for the Raman fields are $\Delta \omega_{\vec{k}_1} = 0$ and $\Delta \omega_{\vec{k}_2} = 0$. A discussion about deviations from these conditions can be found at Ref.~\cite{DLCZ_02}.

Under these assumptions, and with the slow envelope functions written as
\begin{subequations}
\begin{align}
u_r(\vec{r}_i,t) &= q_r(\vec{r}_i)f_r(t) \,,\\
u_w(\vec{r}_i,t) &= q_w(\vec{r}_i)f_w(t) \,,
\end{align}
\end{subequations}
Equation~\eqref{Ai_2} becomes
\begin{align}
&A_i(m_g,m_g) = q_r(\vec{r}_i)q_w(\vec{r}_i)\!\! \sum_{m_s = -F_s}^{F_s} \!\! \frac{d(m_g,ms)}{\hbar^4} \,F(t,z_i)\,, \label{Ai_3}
\end{align}
with
\begin{align}
& F(t,z_i) = \int_0^t \!\!\! dt^{\prime} e^{i(-\Delta_r+a_{ig})t^{\prime}}   \int_0^{t^{\prime}} \!\!\! dt^{\prime\prime}f_r(t^{\prime\prime})e^{i(\Delta_r-a_{is})t^{\prime\prime}} \nonumber \\
& \quad \times \int_0^{t^{\prime\prime}} \!\!\! dt^{\prime\prime\prime} e^{i(-\Delta_w+a_{is})t^{\prime\prime\prime}}  \int_0^{t^{\prime\prime\prime}} \!\!\! dt^{\prime \nu} f_w(t^{\prime\nu})e^{i(\Delta_w-a_{ig})t^{\prime \nu}} . \label{F_1}
\end{align}
Note that the $F$ function depends on the parameters for a specific atom only through $z_i$ that specifies its position along  the quantization axis. In this way, after a certain time, atoms in different parts of the ensemble contribute to the probability amplitude of the process with different phases.

If we consider a uniform distribution of atoms throughout the beam path, and neglecting the $z$ dependence on the $q$ functions, the sum over all atoms may be transformed in the following integral
\begin{align}
\sum_{i=1}^{N} q_r(\vec{r}_i)q_w(\vec{r}_i) \!\!\rightarrow &\frac{N}{V}\int\!\int\!\int dx\,dy\,dz \,q_r(x,y)q_w(x,y) = \nonumber \\ &= \int\!\int dx\,dy\, \frac{q_r(x,y)q_w(x,y)}{A} \frac{N}{L} \int \!dz \nonumber \\ &= \langle q_r(x,y)q_w(x,y) \rangle \,\,N \int_{-L/2}^{L/2} \frac{dz}{L} \,, \label{dist}
\end{align}
where $V= AL$ gives the volume of the excitation region, $A$ its transverse area, and $L$ its length.

Substituting Eqs.~\eqref{Ai_3} and~\eqref{dist} in Eq.~\eqref{p12_5}, we finally obtain
\begin{equation}
p_{12}^{th}(t) = C \left| \sum_{m_g = -F_g}^{F_g} \sum_{m_s = -F_s}^{F_g} \!\!\!D_{m_g} d(m_g,m_s) \int_{-L/2}^{L/2} \!\frac{dz}{L} F(t,z) \right|^2, \label{p12_6}
\end{equation}
where
\begin{equation}
C = N^2 \left| \langle q_r(x,y)q_w(x,y) \rangle \right|^2,
\end{equation}
is a constant. After the read pulse has left the sample (i.e., when $t\rightarrow\infty$), Expression~\eqref{p12_6} is then proportional to the total probability of detecting the pair of photons in one trial. Details on how to compare this expression to the experimental results will be discussed in Sec.~\ref{comparison_theory}. In the experimentally important case of square pulses, it is straightforward to obtain analytical expressions for both $F(t,z)$ and $p_{12}(t)$ in the limit of large $\Delta_w$ and $\Delta_r$.

\subsubsection{Probability density}
\label{density}

Equation~\eqref{p12_6} gives the total probability of detecting one photon in field 2 after detecting a photon in field 1. Now we want to obtain the probability of finding photon 2 between times $t_2$ and $t_2 + \Delta t_2$ and photon 1 between times $t_1$ and $t_1 + \Delta t_1$, for small $\Delta t_2$ and $\Delta t_1$.

The first step in this calculation is to note that Eq.~\eqref{p12_6} can be written as,
\begin{equation}
p_{12}^{th}(t) = |\phi(t)|^2. \label{p12_7}
\end{equation}
The function $\phi(t)$ gives then a probability amplitude for the process where the two photons are found up to time $t$. It consists of an integral over all possible pairs of detection times $(t_2,t_1)$, representing different excitation pathways, and can in principle also be written as
\begin{equation}
\phi(t) = \int_0^tdt_1 \int_{t_1}^tdt_2 P(t_2,t_1)  , \label{phi_2}
\end{equation}
where we considered explicitly $t_2 > t_1$. $P(t_2,t_1)$ represents then a density of probability amplitude.

The probability amplitude for finding photon 2 between times $t_2$ and $t_2 + \Delta t_2$, and photon 1 between times $t_1$ and $t_1 + \Delta t_1$, can be obtained then by restriction over the temporal integral in Eq.~\eqref{p12_6}. Since all the temporal dynamics in Eq.~\eqref{p12_6} is in the function $F(t,z)$, we need to calculate first the restriction of $F(t,z)$ for these specific processes. In order to do so, note that, in the fourth order integral of $F(t,z)$, the emission of photon 2 is described by the last integral (over $t^{\prime}$), while photon 1 emission is described by the third integral (over $t^{\prime\prime\prime}$). The restriction of $F(t,z)$ for the emission of photon 2 between times $t_2$ and $t_2 + \Delta t_2$, and photon 1 between times $t_1$ and $t_1 + \Delta t_1$, is then given by~\cite{restriction}
\begin{align}
G(&t_2,\Delta t_2,t_1,\Delta t_1) = \int_{t_2}^{t_2+\Delta t_2} \!\!\! dt^{\prime} e^{i(-\Delta_r+a_{ig})t^{\prime}} \nonumber \\ 
& \times  \int_0^{t^{\prime}} \!\!\! dt^{\prime\prime}f_r(t^{\prime\prime})e^{i(\Delta_r-a_{is})t^{\prime\prime}} \int_{t_1}^{t_1+\Delta t_1} \!\!\! dt^{\prime\prime\prime} e^{i(-\Delta_w+a_{is})t^{\prime\prime\prime}}   \nonumber \\ 
& \times \int_0^{t^{\prime\prime\prime}} \!\!\! dt^{\prime \nu} f_w(t^{\prime\nu})e^{i(\Delta_w-a_{ig})t^{\prime \nu}}. \label{G_1}
\end{align}

\noindent
Equation~\eqref{G_1} can be directly evaluated for the case of square pulses and large detunings, such that $\Delta_r,\Delta_w >> \Delta t_2^{-1}, \Delta t_1^{-1}$. If the time intervals are also small when compared to the timescale of oscillations determined by the Zeeman shifts (i.e., $\Delta t_2, \Delta t_1 << a_g^{-1}, a_s^{-1}$), then Eq.~\eqref{G_1} can be written as
\begin{equation}
G(t_2,\Delta t_2,t_1,\Delta t_1) = g(t_2,t_1) \Delta t_1\Delta t_2 \,, \label{G_3}
\end{equation}
with
\begin{equation}
g(t_2,t_1) = -\frac{f_r(t_2)f_w(t_1)}{\Delta_r\Delta_w} \, e^{i(a_g - a_s)(t_2-t_1)} \,. \label{g12_1}
\end{equation}
In this case, $F(t,z)$ can be derived by: 
\begin{equation}
F(t,z) = \int_0^t dt_1 \int_{t1}^t dt_2 \,g(t_2,t_1)\;.
\end{equation}
An important remark is that, since any pulse envelope can be approximated by a sum of square pulses of different intensities and small duration, Eq.~\eqref{g12_1} is indeed valid for arbitrary pulse shapes, as long as the envelope temporal variation occurs in a much longer timescale than $\Delta t_1$ or $\Delta t_2$.

The connection between $g(t_2,t_1)$ and the density of probability amplitude $P(t_2,t_1)$ is then made through the relation
\begin{align}
P(t_2,t_1) =& \sqrt{C} \sum_{m_g = -F_g}^{F_g} \sum_{m_s = -F_s}^{F_g} \!\!\!D_{m_g} d(m_g,m_s) \nonumber \\ &\times \int_{-L/2}^{L/2} \frac{dz}{L} \, g(t_2,t_1). \label{P12_1}
\end{align}
Finally, the probability density for detecting one photon from field 1 at time $t_1$ and another from field 2 at $t_2$ is associated to
\begin{equation}
{\cal P}(t_2,t_1)=|P(t_2,t_1)|^2. \label{P_density}
\end{equation}
This is the quantity to be compared with the experimental results of Sec.~\ref{2photon}, for the two-photon wavepacket of the photon pair.

\section{Experiments}
\label{experiments}

Up to now, the experimental implementation of the DLCZ protocol in MOTs have been plagued by extremely short coherence times~\cite{DLCZ_03,DLCZ_04,DLCZ_05,kuzmich2004}. As discussed above, this short coherence time is a result of the action of the MOT quadrupole field over the Zeeman structure of the hyperfine ground states. In the following, we are going to describe a series of experiments that allowed us to obtain photon pairs from the trapped atomic cloud in a situation of very small magnetic field. In this way, we were able to measure coherence times of more than 10 $\mu$s (more than two orders of magnitude longer than the duration of the excitation pulses), and two-photon wavepackets for the photon pairs that do not exhibit distortion by decoherence even when write and read pulses cease overlapping in time~\cite{DLCZ_05}.

The crucial point is to turn off the MOT magnetic field and determine the experimental conditions with a best tradeoff between high repetition rate and high optical density. Note that the atoms fly away from the trap and the density starts to decrease when the magnetic field is turned off. Hence, the MOT field has to be turned off as fast as possible, to decrease the transient time and maximize the region with low magnetic field and high density. A fast turning off of the magnetic field in our metallic vacuum chambers, however, is not straightforward and requires specific techniques, as will be discussed in Sec.~\ref{null}. 

Inside each MOT-off period, it is possible to conduct many trials of the photon pair experiments. These are photon counting measurements that require many events in order to acquire good statistics. Hence, we would like to have as many MOT-off periods as possible to accumulate a large number of trials. However, the MOT needs some time to recover its original density after each off period, and this time limits how often it can be turned off while still keeping a high enough atomic density.

During the process of turning off the magnetic field and determining the proper conditions for the photon counting experiments, it was essential to be able to perform simpler experiments giving direct access to the ground state broadening by the magnetic field. We chose then to setup a copropagating stimulated Raman spectroscopy apparatus to help us in this process. The results for the Raman spectroscopy measurements and the investigation to determine the best experimental conditions for the photon pair generation are described also in Sec.~\ref{null}.     

The nonclassical correlation experiments are discussed in Sec.~\ref{correlation measurements}. There we show that the coherence time increases by more than two orders of magnitude once the magnetic field is switched off, and describe measurements of the shape of the two-photon wavepacket in both situations. In this section, we also compare the experimental results with the theory of Sec.~\ref{decoherence}.

\subsection{Characterization and magnetic field nulling}
\label{null}

\begin{figure}[ht] 
\centerline{\includegraphics[width =7cm,angle=270]{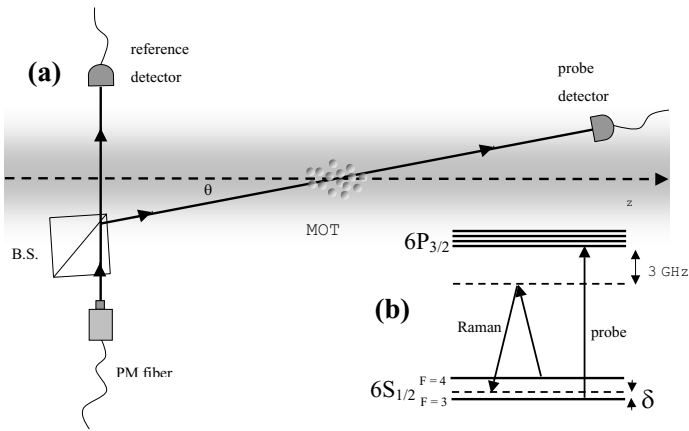}} 
\caption{(a) Experimental Raman spectroscopy setup. The Raman beams and the probe beam are coupled into a polarization maintaining (PM) fiber and sent trough a beam spitter cube (BS). The reflected part is focused into the sample with an angle of 3 degrees with respect to the quadrupole field z axis, while the transmitted part is used as a reference. (b) Relevant level structures and laser frequencies for Raman spectroscopy.} \label{ramansetup}
\end{figure}

As anticipated above, we use copropagating stimulated Raman spectroscopy~\cite{Raman} to  probe directly the broadening of the hyperfine ground states. Our choice for this specific technique is based on the fact that it is insensitive to Doppler broadening, but very sensitive to any broadening caused by magnetic fields, exactly like the spontaneous Raman emission process underlying the photon pair generation in our experiment. Raman stimulated transitions (see Fig.~\ref{ramansetup}a) are two-photon transitions connecting one ground-state hyperfine level to the other one, in which a single photon is absorbed from one Raman beam and another photon is emitted in the other beam by stimulated emission through a virtual level, which is located 3~GHz below the Cesium $D_2$ line in our setup. 

The Raman process is resonant if the frequency difference of the two Raman beams equals the ground-state hyperfine interval, around 9.192631770 GHz for Cesium. In the absence of collisions and transit broadening, this two-photon resonance is very sharp, with a linewidth limited only by the power and duration of the Raman beams~\cite{Raman}. In this way, since the specific value of the hyperfine interval for transitions between $|m_g\rangle$ and $|m_s\rangle$ states changes with the magnetic field, scanning the frequency of one Raman beam with respect to the other gives direct information on the frequency distribution of possible two-photon resonances dislocated by the magnetic field, i.e., on the broadening of the ground state.

Our setup for Raman spectroscopy is shown in Fig.~\ref{ramansetup}a. The two Raman beams and a probe beam are coupled to the same polarization maintaining fiber, which takes the beams close to the MOT and provides good mode-matching between them. The probe beam is coupled with the same polarization as the Raman field connecting the $F=3$ ground state to the virtual level, the other Raman field is coupled with the orthogonal polarization. The lens at the fiber output focus the beam to a diameter of 150~$\mu$m in the MOT region. After the fiber, the beams pass through a 50/50 beam splitter cube. The transmitted parts of the beams are used as a reference to compensate for power fluctuations. The reflected part is directed to the MOT, forming an angle of about $\theta \approx 3^{\circ}$ with the quadrupole-field $z$ axis. The shaded area around the $z$ axis in Fig.~\ref{ramansetup}a indicates the path of one of our trapping beams. The absorption of the probe beam by the atoms in the MOT is then measured with a second detector, by comparing the probe pulse height with MOT on and off.

Before the Raman pulses reach the MOT, an optical pumping cycle moves the whole atomic population to just one of the hyperfine ground states. Note that for the following experiments, we make no attempt to optically pump the atoms onto a specific Zeeman state. Hence, the atomic ensemble is unpolarized and all Zeeman substates are populated. The action of the Raman pulses, of about 150~$\mu$s duration and 10~$\mu$W power, then transfers some population to the initially empty level if their relative detuning matches one of the two-photon transitions of the sample. The probe pulse has a duration of 5~$\mu$s and comes 50~$\mu$s after the Raman pulses. It is resonant with the cycling transition connecting the initially empty ground state to the $6P_{3/2}$ level [$F=4\rightarrow F^{\prime}=5$ if the empty ground state is $F=4$, $F=3\rightarrow F^{\prime}=2$ for empty $F=3$ state]. The probe power is about 50~nW, to guarantee a low saturation of the transition. It is then very sensitive to any change in the initial population, and its absorption indicates that the Raman pulses succeeded in transferring some population from one ground state to the other. 

\begin{figure}[ht] 
\centerline{\includegraphics[width=7cm,angle=270]{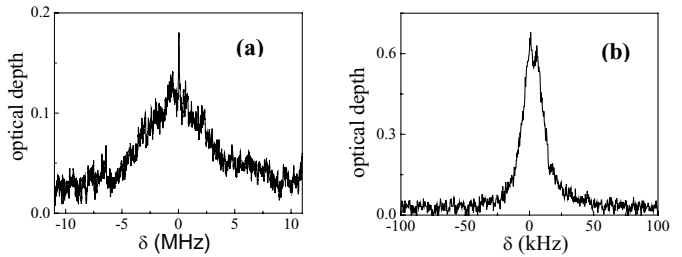}} 
\vspace{-1.5cm}
\caption{(a) Raman trace with the quadrupole MOT magnetic field on. The trace represents the absorbtion of the probe pulse following the Raman beams, as a function of the Raman detuning $\delta$. The line width FWHM is around 5 MHz. (b) Raman trace 4 ms after the quadrupole field has been switched off. The fitted linewidth is 20 kHz, including 10 kHz of power broadening due to the Raman beams} \label{ramanscans}
\end{figure}

In this way, a plot of the medium optical depth for the probe pulse as a function of the detuning between the two Raman fields gives a direct measure of the ensemble distribution of energies in the ground states. Examples of such plots with the MOT magnetic field on and off are shown in Figs.~\ref{ramanscans}a and~\ref{ramanscans}b, respectively. In Fig.~\ref{ramanscans}b the Raman pulses are delayed 4~ms from the moment the magnetic field was turned off, and the nulling of the field was performed using additional bias coils located around the MOT and looking for a reduced width of the Raman trace. From Fig.~\ref{ramanscans}a to~\ref{ramanscans}b, the width of the signal is then reduced by more than two orders of magnitude, from 5~MHz to about 20~kHz. The 20~kHz linewidth of Fig.~\ref{ramanscans}b, however, also includes about 10~kHz that comes from power broadening by the Raman beams. To measure this power broadening, we applied an extra DC field in the $z$ direction in order to split the central peak between the various $m_F \rightarrow m_F^{\prime}$ transitions, and then measured the width of the magnetic-field-insensitive transition $m_F=0 \rightarrow m_F^{\prime}=0$.
\vspace{-1.5cm}   
As mentioned above, the quadrupole field of the MOT should be switched off as fast as possible, in order to maintain the high optical density needed for the DLCZ-type experiments. However, switching off the magnetic field generated by the MOT coils is usually  retarded for two reasons. First, the current in the coils decays exponentially, with a time constant proportional to the inductance of the coils. Second, the field decay time is increased by eddy currents in the metallic part of our vacuum chamber. Depending on the metallic configuration of chamber and coils, the transient period can last for tens of ms. In order to obtain a faster transient, we use a fast-switching electronic circuit~\cite{boff_01,boff_02}. This circuit allows a quick reversal of the current in the quadrupole coils in order to compensate for the eddy currents, and resulted in a substantial reduction of the transient time in our system.

\begin{figure}[ht]
\centerline{\includegraphics[width =6.7 cm,angle=270]{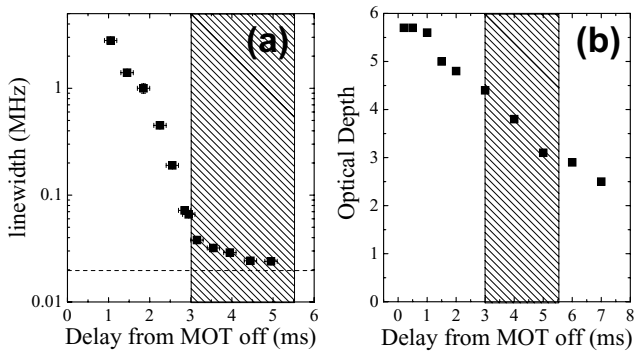}}
\vspace{-1.0cm}
\caption{(a) Evolution of the ground state linewidth and (b) of the optical depth of the sample as a function of the delay from the time when the current is switched off in the MOT coils. The linewidth is measured with Raman spectroscopy. The dashed line represents the measured power broadening due to the Raman beams. The OD is determined by measuring the absorbtion of a probe pulse in the sample. In both graphs, the dashed area represents the window used for measuring correlations at the single photon level.} \label{linewidth}
\end{figure}

A detailed description of the magnetic field transient is given in Fig.~\ref{linewidth}a, which plots the Raman scan linewidth as a function of the delay from the moment the field was switched off. Figure~\ref{linewidth}a then shows the timescale over which the ground state has its energy-distribution profile changed from Fig.~\ref{ramanscans}a to Fig.~\ref{ramanscans}b. We can see that after a few miliseconds, the linewidth asymptotically reaches a plateau, given by the residual DC field in the chamber, that we estimate in this case to be on the order of 10 mG. The dashed line in Fig.~\ref{linewidth}a indicates the measured power broadening. Shorter transients can be obtained with a different metallic chamber configuration (like in Ref.~\onlinecite{boff_01}) or using non-metallic vacuum chambers.

In order to estimate the optimal region for photon counting measurements, it is important to independently measure the decay of the optical depth after the magnetic field is switched off. In our setup this is done in a straightforward way by turning off the Raman beams and using a probe pulse close to resonance with the ground state that concentrates all the atomic population. The results of such measurement are shown in Fig.~\ref{linewidth}b, for which the population was initially pumped to $F=4$ and the probe tuned 10~MHz below the $F=4\rightarrow F^{\prime}=5$ transition. The optical depth measurements in Fig.~\ref{linewidth}b were obtained from the absorption at 10~MHz detuning and assuming a Lorentzian lineshape for the atomic transition with a natural linewidth corrected for power broadening by the probe beam.   

Together, the results in Figs.~\ref{linewidth}a and~\ref{linewidth}b allow us to determine an optimal window for the experiments of Sec.~\ref{correlation measurements}, i.e., between 3 and 5.5~ms (dashed region in both figures). The lower limit of this region is determined by the moment when the residual magnetic field reaches a reasonably small value corresponding to an acceptable decoherence time, and the higher limit by the restriction that the density should not vary too much throughout the region. We accepted a variation of about 30\% in the density. The linewidth varies by about 30~kHz in the same interval.

\begin{figure}[ht]
\centerline{\includegraphics[width =8.5 cm,angle=0]{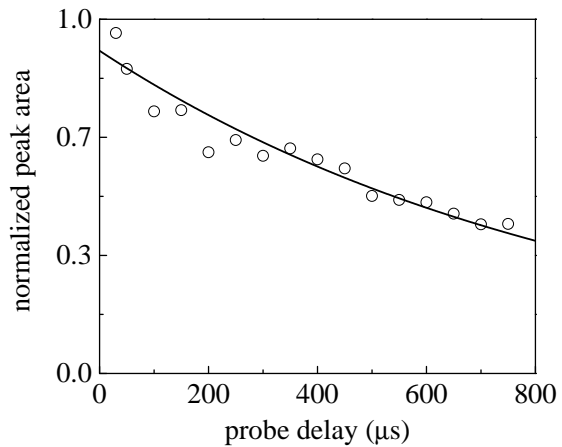}}
\vspace{-1.7cm}
\caption{Diffusion of atoms out of the excitation region. The
solid line is an exponential fit with a time constant of 900 $\mu
s$. The Raman beam diameter is 150 $\mu$m.} \label{diffusion}
\end{figure}

A better cancellation of the magnetic field can in principle lead to even smaller linewidths and, consequently, longer coherence times. However, improvements along this line will eventually be limited by a different problem: the diffusion of atoms out of the excitation region. This effect of course depends on the temperature of the sample and on the diameter of the excitation beams. In order to directly measure this diffusion time, we use again Raman spectroscopy. In this case, Raman traces are recorded as a function of the delay between the Raman pulses and probe. The measurement is done when the magnetic field is off, such that there is only one narrow peak in the Raman trace, like in Fig.~\ref{ramansetup}d. In this case, the area of the peak profile is proportional to the number of atoms in the excitation region. Figure~\ref{diffusion} shows a plot of this area as a function of delay. We see that the population decays with a time constant of 900 $\mu s$, as given by an exponential fit to the data (solid line). Note that this measurement was done with beams that have 150 $\mu$m diameter, while in the correlation measurements described later we use beams with 60 $\mu$m diameter, leading to a diffusion time of the order of 360 $\mu$s. 

\vspace{1.4cm}

\subsection{Nonclassical correlations} \label{correlation
measurements} 

In order to characterize the coherence time of the system for various quantum information applications, e.g. for the DLCZ protocol or for generation of conditional single photons, the measurements must be performed at the single-photon level. In particular, one must know how long a single excitation can be stored in the quantum memory. For this purpose, we perform correlation measurements between fields $1$ and $2$ as a function of the time delay $\Delta t$ between write and read pulses, thereby probing how the nonclassical character of these correlations ( and hence of the correlations between field 1 and the collective atomic excitations) is preserved during the storage process. 

In order to investigate the quantum nature of the correlations, we use the fact that there exists a well-defined border between the classical and quantum domains for fields 1 and 2 that can be operationally accessed via coincidence detection, as was first demonstrated in the pioneering work by Clauser~\cite{Clauser}. In this way, we measure the joint
detection probability $p_{12}$ for detecting a photon in both fields 1 and 2 in the same trial, and the probabilities $p_1$ and $p_2$ to register a single detection event in field 1 and field 2, respectively. By splitting field $i$ with a 50-50 beamsplitter and directing the output to the two detectors, the joint probabilities $p_{ii}$ are also measured, where $i= 1$ or 2. Fields for which the Glauber-Sudarshan phase-space function is well-behaved (i.e., classical fields) are constrained by a Cauchy-Schwarz inequality for the various probabilities~\cite{Clauser,MandelBook}, namely:
\begin{equation}
R =
\frac{[g_{12}(t)]^2}{g_{11}\,g_{22}}
\leq 1 \,, \label{CSineq}
\end{equation}
where $g_{11}\equiv p_{11}/p_1^2$, $g_{22}\equiv p_{22}/p_2^2$, $g_{12}(t)\equiv p_{12}/(p_1 p_2)$, and $t$ denotes the time separation between the detection of photons 1 and 2. In our system, $g_{11}=g_{22}=2$ in the ideal case. However, in practice, $g_{11}$ and $g_{22}$ are measured to be smaller than 2, due to various experimental imperfections. Hence in our case measuring $g_{12}>2$ heralds nonclassical correlations, and in the following we will use this quantity as another figure of merit to quantify the loss of coherence in the quantum memory.

\begin{figure}[ht]
\centerline{\includegraphics[width =8 cm,angle=0]{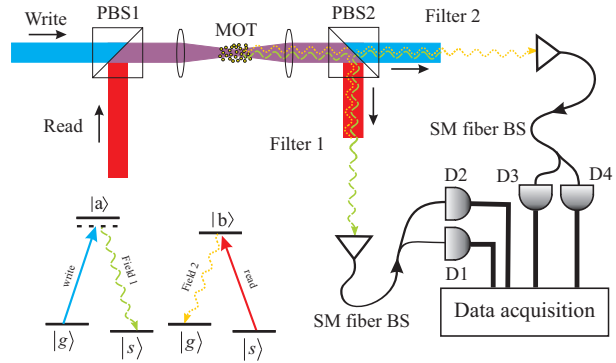}}
\caption{Experimental setup. Write and read pulses propagate sequentially into a cloud
of cold Cs atoms (MOT), generating pairs of correlated output photons 1 and 2. The
write and read pulses have orthogonal polarizations, are combined 
at polarizing beam splitter PBS1, and then focused in the Cs MOT with a waist of
approximately 30 $\mu m$. The output fields are split by PBS2, which also serves as a first
stage of filtering the (write, read) beams from the (1,2) fields. For example, field 2 is
transmitted by PBS2 to be subsequently registered by detector D3 or D4 while the read pulse
itself is reflected at PBS2. Further filtering is achieved by passing each of the
outputs from PBS2 through separate frequency filters. SM stands for single mode.} \label{setupcorr}
\end{figure}

The experimental setup used to measure nonclassical correlations between fields $1$ and $2$ is shown in Fig. \ref{setupcorr}. As already mentioned the sample consists in a cold atomic ensemble of Cesium atoms in a magneto-optical trap. Each trial consists of a period of cooling and trapping, and of a  period of measurement during which all the beams responsible for cooling and trapping the atoms are switched off. During the measurement period, the atoms are initially prepared in level $|g\rangle$ (F=4) by optical pumping  with a laser beam resonant with the transition $6S_{1/2}(F=3) \rightarrow 6P_{3/2}(F'=4)$. 

A laser pulse with 150 ns duration from the write beam then illuminates the sample. The write beam is tuned near the $|g\rangle \rightarrow |a\rangle$ (corresponding to $F=4\rightarrow F'=4$ of the $D_2$ line, at 852 nm) and induces spontaneous Raman scattering to the initially empty level $|s\rangle$ ($F=3)$. The intensity of the pulse is made sufficiently weak, such that the probability of creating more than one excitation in the symmetric collective mode is very low. After a variable delay $\Delta t$, the stored excitation is converted into a photon in field $2$, by sending a read pulse tuned to the transition $|s\rangle \rightarrow |b\rangle$ (corresponding to $F=3 \rightarrow F'=4$ transition of the $D_1$ line, at 894 nm). The write and read beams are orthogonally polarized and combined at the polarizing beam splitter PBS 1 (see Fig. \ref{setupcorr}). At PBS 1, the write and read beams are spatially mode-matched with a measured overlap of about $93 \%$. The beams are focussed to a waist of about $30 \mu m$ in the sample region.

After the MOT, fields 1 and 2 are detected at the two different outputs of PBS 2. A challenging aspect of the experiment is to separate the classical pulses from the weak nonclassical fields, since they are temporally and spatially overlapped, and their frequencies are only 9~GHz apart. This is done in several steps, which are explained in detail in Refs.~\cite{DLCZ_03},~\cite{DLCZ_04}, and~\cite{DLCZ_05}. After the filters, fields $1$ and $2$ are coupled into optical fibers, split by 50/50 fiber beam splitters, and detected by four single-photon Silicon avalanche photodiodes (APD). Finally, the electronic signals of the APDs are sent to a data acquisition card, in order to record the detection events and analyze the correlations.

\subsubsection{Coherence time measurements}
\label{comparison_theory}

\begin{figure}[ht]
\centerline{\includegraphics[width=8.5cm]{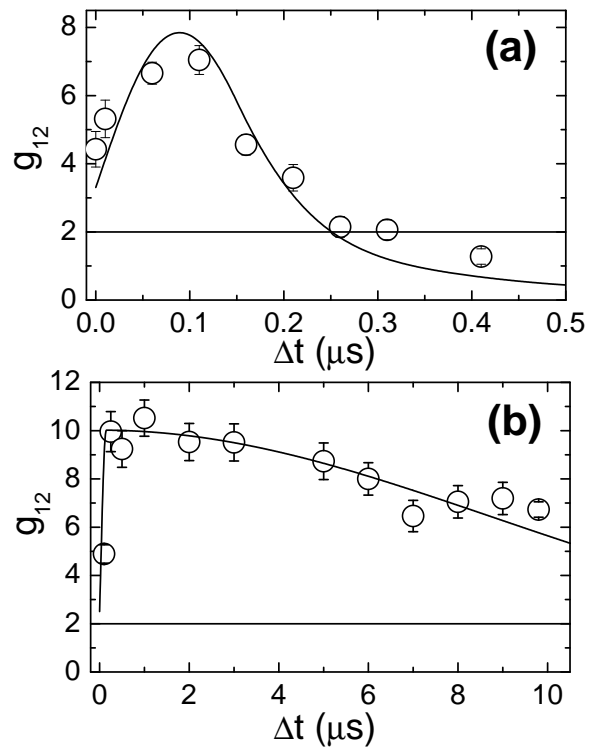}}
\vspace{-0.7cm}
\caption{ Measurement of $g_{12}$ as a function of the storage time, (a) with the quadrupole field on (taken from \cite{DLCZ_05}) and (b) with the quadrupole field off. The observed decay in (b) is consistent with the residual magnetic field in the chamber, as measured by Raman spectroscopy.} \label{g12old}
\end{figure}

In order to characterize the system's coherence time, we measure $g_{12}$ and $R$ as a function of the delay $\Delta t$ between write and read pulses. We then compare the theoretical quantity $\tilde{p}_{12}(\Delta t) = \xi p_{12}^{th}(\Delta t)$ to the measured $g_{12}(\Delta t)$ by way of a single overall scaling parameter $\xi$ for all $\Delta t$, as the rate of single counts in fields 1 and 2 ($p_1$ and $p_2$) is measured not to depend on $\Delta t$, to within 20\%. In Fig.~\ref{g12old}a we show our results for $g_{12}$ with the MOT magnetic field on together with the corresponding theoretical fitting. This figure was presented already in a previous article~\cite{DLCZ_05} and shows a fast decay of the coherence between fields 1 and 2, taking place in a time scale of less than 200~ns. Note, however, that the coherence time is actually smaller than 100~ns, since the write pulse itself has a duration of 150~ns. The repetition rate of the trials in this case is 250~kHz. The rate of coincidence events (detection of photon 1 and photon 2 within the same trial) is between 2 and 3 counts per second.

The theoretical joint probability $p_{12}^{th}$ is calculated from Eq.~\eqref{p12_6}, assuming $C=1$. In this way, we need to perform integrals of the $F$ function over the $z$ coordinate. This function depends on $z$ only through the parameters $a_g$ and $a_s$. The atomic ensemble is assumed to be initially unpolarized, i.e., with the atoms evenly distributed among all Zeeman states of the $|g\rangle$ level. For the ground states of Cesium, we have that the hyperfine Land\'e factors $g_g$ and $g_s$ of levels $|g\rangle$ and $|s\rangle$, respectively, are given by $\mu_B g_g/h = - \mu_B g_s/h = 0.35\,{\rm MHz/G}$, so that we can write
\begin{subequations}
\begin{align}
a_g &= 2\pi K m_g \left( \frac{z}{L} \right)\,, \\
a_s &= -2\pi K m_s \left( \frac{z}{L} \right)\,, 
\end{align}
\label{as}
\end{subequations}

\noindent
where we considered the magnetic field for the MOT in the form $B_z = b z$, with $b$ the field gradient in the center of the MOT, and the constant $K$ given by
\begin{equation}
K = \frac{\mu_B g_g b L}{h}\,.
\end{equation}
The value of $K m_F$ gives an estimate for the inhomogeneous broadening associated with level $|F,m_F\rangle$ due to the magnetic-field gradient $b$. Note that writing $a_g$ and $a_s$ as in Eqs.~\eqref{as} allows us to perform all spatial integrations over the dimensionless coordinate $s = z/L$, and to combine many of the relevant experimental parameters in a single parameter ($K$). For our experiment, $L = 3.6$~mm and $b = 8.7$~G/cm, so that $K = 1.1$~ MHz. This $K$ value is  consistent with the measurement of the ground-state broadening shown in Fig.~\ref{ramanscans}a. 

The solid curve in Fig.~\ref{g12old}a shows the theoretical fitting of $\tilde{p}_{12}(\Delta t)$ to the experimental data. We considered $K = 1.1$~MHz in the theory, as estimated above for our experimental conditions. The only fitting parameter used was $\xi$, which was found to be $\xi = 1.05 \times 10^8$. Note that the theoretical quantity $p_{12}^{th}$ gives the probability for joint detection of the two photons, while $g_{12}$ is a measure of this joint probability normalized by the probability of uncorrelated coincidence detections. Thus the scaling factor $\xi$ should be given roughly by the inverse of the probability for these uncorrelated coincidences. A theoretical estimation for this value is given by $\xi^{th} = [p_{12}^{th}(\Delta t \rightarrow \infty)]^{-1}$, i.e., the inverse of the theoretical joint probability after the coherence has completely decayed. For the solid curve in Fig.~\ref{g12old}a, we find $\xi^{th} = 1.96 \times 10^8$. The difference between $\xi$ and $\xi^{th}$ can be attributed to other sources of uncorrelated coincidences (such as dark counts in the detectors, or leakage from the filters) that are not accounted by the theory, which leads to $\xi < \xi^{th}$. It is also important to have in mind that the noise floor is higher when the pulses are overlapping, since there is more leakage from the filters in this condition. This results in some extra discrepancy when comparing theory to experiment by means of one single scaling parameter to all regions of Fig.~\ref{g12old}a.

\vspace{-0.3cm}

The $g_{12}(\Delta t)$ measurements with magnetic field off are presented in Fig.~\ref{g12old}b. In this case, we use the information acquired from the investigation of Sec.~\ref{null} and turn off the field for a duration of 5.5~ms, at 40~Hz repetition rate. From the magnetic-field-off period, we use for correlation measurements only the 2.5~ms window shown in Fig.~\ref{linewidth}. This 2.5~ms window is then divided in 208 trial periods of 12~$\mu$s, which results in an overall repetition rate of 8.3~kHz. In the beginning of each trial, the trap light of the MOT (tuned in the $F=4$ to $F^{\prime}=5$ transition of the $D_2$ line) is turned on for 0.6~$\mu$s, and its repumper laser (tuned from $F=3$ to $F^{\prime}=4$) for 1~$\mu$s. This procedure prepares the system in the proper initial state, with all atoms at the $F=4$ hyperfine level of the ground state. In this case, the rate of coincidence counts drops to about 0.33 coincidences/s.

Figure~\ref{g12old}b shows then an increase of more than two orders of magnitude on the coherence time of the system, when the magnetic field is turned off. The coherence time is now limited mainly by the rate at which we can turn off the magnetic field, and also to some extent by our ability to magnetically isolate the system. Note that in Fig.~\ref{linewidth}a the Raman-trace linewidth indicates that the magnetic field in the measurement window is still decaying. The solid curve in Fig.~\ref{g12old}b gives the decay theoretically expected for a magnetic-field gradient such that $K = 12$~kHz, corresponding to magnetic fields of the order or smaller than $30$~mG acting on the ensemble. This gives a reasonable approximation to the behavior of $g_{12}$ under the action of the residual magnetic field, even though the spatial dependence of this field can be more complicated than a simple linear gradient. The change in K from 1.1 MHz to 12 kHz is consistent with the reduction of the ground state linewidth between the two cases, as measured directly by the Raman spectroscopy setup. Finally, for Fig.~\ref{g12old}b $\xi = 0.67 \times 10^8$ and $\xi^{th} = 2.2 \times 10^8$. 

From Fig. \ref{g12old}.b, we see that the correlations are still highly nonclassical after a storage time of 10 $\mu s$. However, from the theoretical fitting we can infer that $g_{12}$ should became smaller than 2 at about 25~$\mu$s, which gives an estimation for our quantum memory time.  

\begin{figure}[ht]
\centerline{\includegraphics[width=8.5cm]{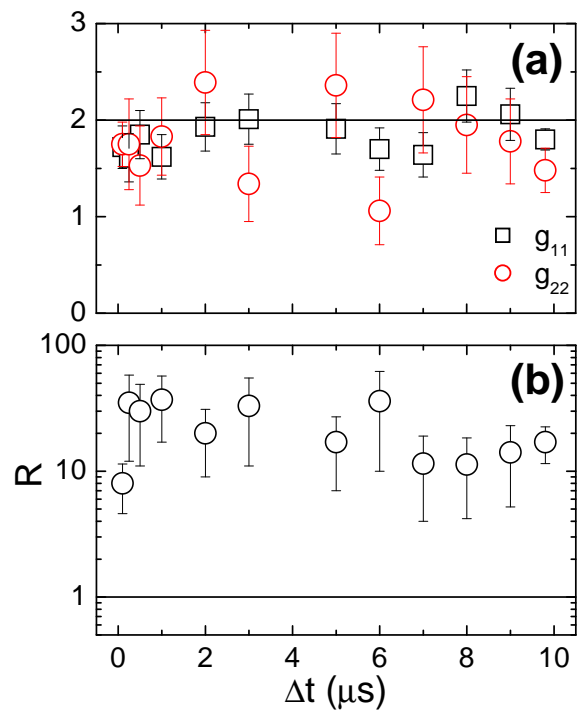}}
\vspace{-0.8cm}
\caption{(a) Measurement of $g_{11}$ (open squares) and $g_{22}$ (open circles) as a function of the storage time. (b) Measurement of the coefficient R as a function of the storage time. The big statistical errors are mainly due to statistical uncertainties in the measurement of $g_{11}$ and $g_{22}$. The points at 10$\mu$s have been measured for a much longer time and exhibit smaller statistical error. } \label{Rexper}
\end{figure}

As discussed above, the  measurements with $g_{12}>2$ give a strong indication of the nonclassical correlations observed in our system, based on reasonable assumptions for $g_{11}$ and $g_{22}$. The most appropriate verification of the nonclassical nature of fields 1 and 2, however, is given by the measurement of $R$ as defined in Eq.~\eqref{CSineq}. Such measurements with the magnetic field off are shown in Fig.~\ref{Rexper}. More specifically, in Fig.~\ref{Rexper}a we show the measurements of $g_{11}$ and $g_{22}$ for the same data points of Fig.~\ref{g12old}b. Substituting the results of Figs.~\ref{g12old}b and~\ref{Rexper}a in~\eqref{CSineq}, we then obtain the values of $R$ shown in Fig.~\ref{Rexper}b, which confirm the strong nonclassical correlation present in our system for more than 10~$\mu$s. 

The $R$ measurement presents considerably larger error bars than for $g_{12}$. This comes from the large statistical uncertainties involved in the determination of $g_{22}$, which requires measurement of the two-photon component of field 2~\cite{DLCZ_04}. For this reason, we decided to carry out a much longer run of the experiment for the longest coherence time we were able to probe, 10~$\mu$s, which resulted in the considerably smaller statistical error of this point.  

\subsubsection{Two-photon wavepackets}
\label{2photon}

Central to the DLCZ protocol is the ability to write and read collective spin excitations into and out of an atomic ensemble, with efficient conversion of discrete spin excitations to single-photon wavepackets. A critical aspect of such wave packets is that they are emitted into well defined spatiotemporal modes to enable quantum interference between emissions from separate ensembles (e.g., for entanglement based quantum cryptography~\cite{DLCZ_01}). 

The high efficiencies achieved in the work of Ref.~\onlinecite{DLCZ_04} enabled us to investigate in detail the temporal properties of the nonclassical correlations between emitted photon pairs~\cite{DLCZ_05}, providing a direct look at various important features of the two-photon wavepacket (field 1 + field 2) generated by the system. In the following analysis, our main quantity of interest is $p_{\tau}(t_{1},t_{2})$, the joint probability for photoelectric detection of photon 1 at time $t_1$ and photon 2 at time $t_2$ within a time window of duration $\tau$. The times for this quantity are counted starting from the beginning of the write pulse. This quantities is determined from the record of time-stamped detections on all four photodetectors. The detectors have a time resolution of 2~ns (minimum bin size), but usually we need to consider larger bins to acquire enough events for the statistics.

\vspace{-0.5cm}

\begin{figure}[htb]
    \begin{center}
        \includegraphics[width = 8.5cm, angle=0]{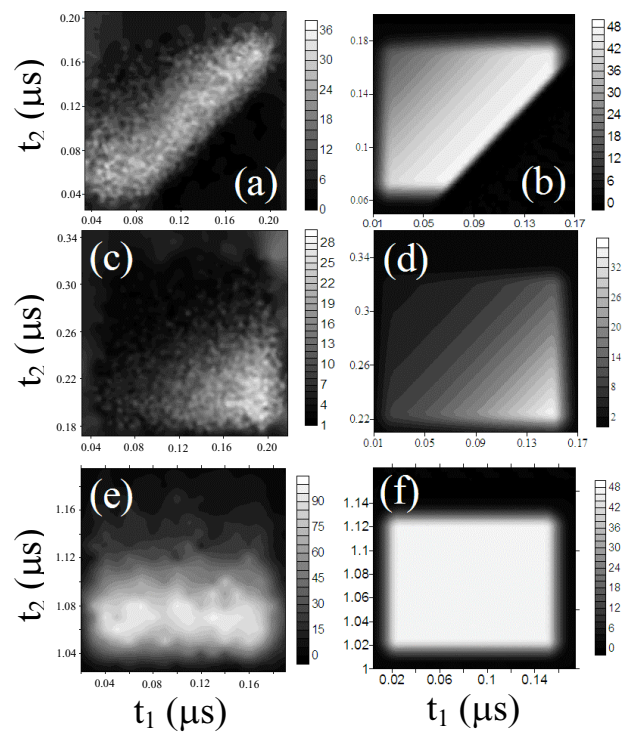}
    \end{center}
    \vspace{-1.0cm}
    \caption{Theory and experiment for two-photon wavepackets $P_\tau(t_1,t_2)$. (a) Measured two-photon wavepackets for the case where write and read pulses are overlaped with a delay of 50 ns, with the quadrupole magnetic field on. (b) Theoretical predictions for the same conditions as in (a). (c) Measured two-photon wavepackets for the case of consecutive (non overlapping) write and read puses with a delay of 200 ns, with quadrupole field on. (d) Theoretical predictions for the same conditions as in (c). (e) Measured two-photon wavepackets for nonoverlapping write and read pulses, with quadrupole field off. The delay between write and read pulses is 1 $\mu$s. (f) Theoretical  predictions for the same conditions as in (e). The vertical scales are given in arbitrary units proportional to the joint probability of detecting photons 1 and 2. See text for further details.}
    \label{fig:overlap}
\end{figure}

In our earlier experiments~\cite{DLCZ_05}, we focused on two cases: (\textit{I}) nearly simultaneous application of write and read pulses with offset $\Delta t=50$~ns shorter than the duration of either pulse, and (\textit{II}) consecutive (non overlapping) application of write and read pulses with $\Delta t=200$~ ns. Results for $p_{\tau}(t_{1},t_{2})$ are presented in Fig.~\ref{fig:overlap}. In case (\textit{I}), Fig.~\ref{fig:overlap}a shows that $p_{\tau }(t_{1},t_{2})$ peaks along the line $t_{2}-t_{1}=\delta t_{12}\simeq 50$ ns with a width $\Delta t_{12}\simeq 60$ ns, in correspondence to the delay $\delta t_{12}$ and duration $\Delta t_{12}$ for read-out associated with the transition $|s\rangle \rightarrow |b\rangle \rightarrow |g\rangle $ given an initial transition $|g\rangle \rightarrow |a\rangle \rightarrow |s\rangle$~\cite{DLCZ_06}. In case (\textit{II}) with the read pulse launched $200$ ns after the write pulse, the excitation is ``stored" in the atomic ensemble until the readout. The production of correlated photon pairs should now be
distributed along $t_{2}\simeq \Delta t + \delta t_{12}$ with width $\simeq \Delta t_{12}$. Instead, as shown in Fig.~\ref{fig:overlap}c,
$p_{\tau }(t_{1},t_{2})$ peaks towards the end of the write pulse (i.e., $t_{1}\gtrsim 100$ ns), and near the
beginning of the read pulse (i.e., $200\lesssim t_{2}\lesssim 300$ ns). Early events for field $1$ lead to fewer
correlated events for field $2$, as $p_{\tau }(t_{1},t_{2})$ decays rapidly beyond the line $t_{2}-t_{1}=\tau _{d}\simeq 175$ ns. The marked contrast between $p_{\tau }(t_{1},t_{2})$ for $\Delta t=50$ and $200$ ns results in a diminished ability for the conditional generation of single photons from excitation stored within the atomic ensemble \cite{DLCZ_04} and, more generally, for the implementation of the DLCZ protocol for increasing $\Delta t$. The underlying mechanism is again decoherence within the ensemble. 

By contrast, when the magnetic field is turned off, this distortion in the two-photon wavepacket is eliminated due to the extended coherence time. We now observe the shape shown in Fig.~\ref{fig:overlap}e. The delay in Fig.~\ref{fig:overlap}e is $\Delta t = 1 \; \mu$s.

The theoretical results corresponding to these three situations are shown in frames (b), (d), and (f) of Fig.~\ref{fig:overlap}. These are plots of Eq.~\eqref{P_density} averaged over 4~ns time windows for both $t_2$ and $t_1$, the same time window used for the experimental data. We also considered pulses of trapezoidal shape, with 20~ns  rising time, and FWHM of 150~ns for the write pulse and 120~ns for the read pulse. These values correspond to the experimental parameters. The only effect of both the time window and pulse rising time is to smooth the edges of the distribution. Differently from the case of integrated probabilities, it is necessary here to introduce more details in the description of the pulse shapes, since the theoretical description for this signal predicts that it is directly related to the pulse profiles [see Eq.~\eqref{g12_1}].
    
The main point that calls our attention in these figures is the fact that the theory offers a reasonable explanation for the data from consecutive pulses ($\Delta t = 200$~ns) with magnetic field on, but not for overlapping pulses or $\Delta t = 1\;\mu$s with magnetic field off. This discrepancy can be simply understood, however, if we remember that one of the main approximations of our theory is to consider low intensities for both write and read pulses. At low intensities and zero magnetic field, the theory gives a small and constant probability for the photon 2 emission after photon 1. From Eq.~\eqref{g12_1}, we see that the magnetic field introduces different phases for different groups of atoms. These different phases are proportional to the time difference between the emission of photons 2 and 1, and result in an overall decay of the probability of emission of the second photon over time. In Figs.~\ref{fig:overlap}b and~~\ref{fig:overlap}f, however, we see that the predicted decay time is much longer than the one inferred from the experimental data.

On the other hand, for the actual experiment, the high intensity of the read pulse should lead to a fast emission of photon 2 once the atom is transferred to level $F_s$. This is consistent with the short duration of correlation $\Delta t_{1,2}$ in Figs.~\ref{fig:overlap}a and~\ref{fig:overlap}e, which can be understood as coming from the fast depletion of the $F_s$ state. However, this reasoning cannot explain the shape of Fig.~\ref{fig:overlap}c, since the strong excitation alone should result in a similar fast depletion in the beginning of the read pulse for any detection time of photon 1 (as seen in Fig.~~\ref{fig:overlap}e). The good comparison between Figs.~\ref{fig:overlap}c and~\ref{fig:overlap}d comes from the fact that the decay due to the magnetic field takes place before the delayed readout process occurs. The shape in Fig.~\ref{fig:overlap}c is then a convolution of a uniform excitation probability over $t_1$ (like in Fig.~\ref{fig:overlap}e) with the excitation-probability distribution of~\ref{fig:overlap}d. 

\vspace{-0.5cm}

\section{Optical pumping}
\label{proposals}

\begin{figure}[ht]
\centerline{\includegraphics[width =8.5 cm]{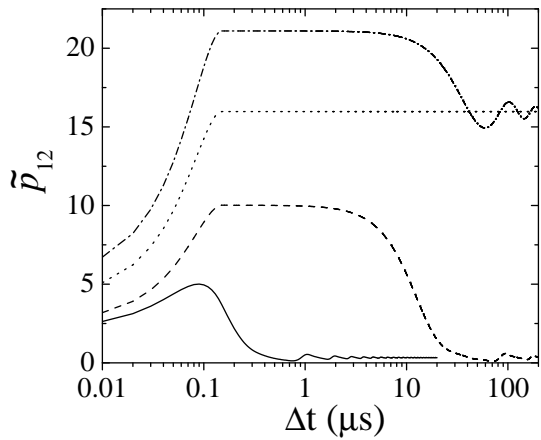}}
\vspace{-1.7cm}
\caption{Variation of $\tilde{p}_{1,2}$ with the delay $\Delta t$ between write and read pulses for (solid curve) $K=1.1$~MHz and an unpolarized sample, (dashed curve) $K=12$~kHz and an unpolarized sample, and (dash-dotted curve) $K=12$~kHz and an initially spin polarized sample with all atoms in $|F=4, m_F = 0\rangle$. The dotted curve corresponds to an initially spin polarized sample classically excited by fields with polarizations such that only a magnetic insensitive transition is allowed, see text for details. The same arbitrary scaling factor was used for all curves.} \label{op}
\end{figure}

The theory developed to explain the data in Fig.~\ref{g12old} can also be used to devise new ways to improve the system. The inclusion of the Zeeman structure in the theory, for example, allows the study of different polarization schemes for both classical excitation and photon detection. It also allows the investigation of the role of the atomic initial state on the measured correlations. In Fig.~\ref{op} we give two examples of possible ways to improve the system. The solid and dashed lines in the figure represent the two experimental conditions of Fig.~\ref{g12old} (initially unpolarized samples with $K=1.1$~MHz and $K=12$~kHz), but now with the same scaling factor. The dash-dotted curve shows how the $K=12$~kHz curve changes if the system is initially spin polarized, with all atoms in the $|F=4, m_F = 0\rangle$ state. Note that in this case the value of $\tilde{p}_{1,2}$ considerably increases, and the system develops a plateau coming from the predominant transition $|F=4, m_F = 0\rangle \rightarrow |F=3, m_F = 0\rangle \rightarrow |F=4, m_F = 0\rangle$, which is magnetic-field insensitive. Furthermore, it is possible to devise a polarization scheme of excitation that allows only this specific transition for any $\Delta t$, e.g. as  when the write pulse and field-1 detection are $\sigma^+$ polarized, and the read pulse and field-2 detection are $\sigma^-$. This is the case for the dotted curve in Fig.~\ref{op}.

\vspace{1.4cm}

The idealized improvements described by the dotted and dash-dotted curves of Fig.~\ref{op}, however, will probably be limited by two effects which are not taken into account by the theory. First, in our experimental setup we should see a decay with a timescale on the order of 360~$\mu$s due to the average time the cold atoms take to cross the 60~$\mu$m beam diameter of the classical write and read pulses. Second, the theory assumes the presence of a magnetic field predominantly in the z direction, which defines the quantization axis. This can be obtained by applying an extra DC magnetic field along that direction,\cite{OptPump_01,OptPump_02} but any residual transverse field should lead to some decay of the plateau. In spite of these restrictions, however, we believe that such improvements could lead to an increase of more than an order of magnitude over the largest experimental decoherence time of Fig.~\ref{g12old}. It is also clear that there is a benefit in the careful preparation of the initial state for the magnitude of the measured correlations. This is an important point that should also be taken into account when considering the implementation of the DLCZ protocol in vapor cells. 

\section{Conclusion}
\label{conclusion}
We have presented a detailed study of the decoherence processes in the generation of photon pairs from atomic ensembles, via the DLCZ protocol of ref. \cite{DLCZ_01}. We have identified the main cause of decoherence for cold atoms in magneto-optical traps as being the inhomogeneous broadening of the hyperfine ground states due to the quadrupole magnetic field used to trap the atoms. A detailed theory has been developed to model this effect. We also reported a series of measurement to characterize and control the decoherence using copropagating stimulated Raman scattering. These measurement allowed us to switch off the quadrupole magnetic field in a controlled way. With the magnetic field off, we observed highly nonclassical correlations between the two emitted photons, for a storage time of up to 10 $\mu$s, an improvement of more than two orders of magnitude compared to previous results with cold atoms. Furthermore, contrary to all related experiments reported up to now, the coherence time is now two orders of magnitude larger than the excitation pulses duration. This is a crucial step in order to use atomic ensembles as a quantum memory to store conditional single photon states or entanglement between two distant ensembles. 
\section*{Acknowledgments}

This work is supported by ARDA, by the Caltech MURI Center for Quantum Networks, and by the NSF. D.F. acknowledges financial support by CNPq (Brazilian agency). H.d.R. aknowledges financial support by the Swiss National Science Foundation.


\begin{thebibliography}{2}
\bibitem{restriction}
Equation~\eqref{G_1} can also be deduced in a more formal way, beginning by considering that the probability amplitude for detecting photon 2 between $t_2$ and $t_2+\Delta t_2$ is given by $\phi(t_2 + \Delta t_2) - \phi(t_2)$, then following all the way down by similar restrictions, and finally making the approximation of small time intervals.
\end{thebibliography}
\end{document}